\documentclass[conference,a4paper]{IEEEtran}
\IEEEoverridecommandlockouts

\usepackage[dvips]{graphicx}  
\usepackage{url}
\usepackage{fancyhdr}

\begin{document}
\title{Analysis of the Social Community\\ Based on the Network Growing Model\\ in Open Source Software Community
\thanks{\copyright 2015 IEEE. Personal use of this material is permitted. Permission from IEEE must be obtained for all other uses, in any current or future media, including reprinting/republishing this material for advertising or promotional purposes, creating new collective works, for resale or redistribution to servers or lists, or reuse of any copyrighted component of this work in other works.}
}

\author{
\IEEEauthorblockN{Takumi Ichimura}
\IEEEauthorblockA{Department of Management and Systems,\\
Prefectural University of Hiroshima.\\
1-1-71, Ujina-Higashi, Minami-ku, \\
Hiroshima 734-8558, Japan.\\
E-mail: ichimura@pu-hiroshima.ac.jp}
\and
\IEEEauthorblockN{Takuya Uemoto}
\IEEEauthorblockA{Graduate School of Comprehensive Scientific Research,\\
Prefectural University of Hiroshima.\\
1-1-71, Ujina-Higashi, Minami-ku, \\
Hiroshima 734-8558, Japan.\\
E-mail: yslius7221@gmail.com}
}

\maketitle

\pagestyle{fancy}{
\fancyhf{}
\fancyfoot[R]{}}
\renewcommand{\headrulewidth}{0pt}
\renewcommand{\footrulewidth}{0pt}

\begin{abstract}
The social community in open source software developers has a complex network structure. The network structure represents the relations between the project and the engineer in the software developer's community. A project forms some teams which consist of engineers categorized into some task group. Source Forge is well known to be one of open source websites. The node and arc in the network structure means the engineer and their connection among engineers in the Source Forge. In the previous study, we found the growing process of project becomes strong according to the number of developers joining into the project. In the growing phase, we found some characteristic patterns between the number of agents and the produced  projects. By such observations, we developed a simulation model of performing the growing process of project. In this paper, we introduced the altruism behavior as shown in the Army Ant model into the software developer's simulation model. The efficiency of the software developing process was investigated by some experimental simulation results.
\end{abstract}

\begin{IEEEkeywords}
Social network, Open source software, Project growing process, Altruism simulation
\end{IEEEkeywords}

\IEEEpeerreviewmaketitle

\section{Introduction}
\label{sec:Introduction}
A social community is composed of a group of individuals with the common interest and purpose. The size of community is determined by the number of people and their activities. We can consider that the attraction to the community increased according to the integration of the interaction of the personal behaviors, and then it grows to be a larger society.

In the community of open source software (OSS) developers and projects, we can observe the collaborative social network in the developing process of system. Madey et al.\cite{Madey03} has been developed the structural and the dynamic mechanisms that govern the topology and evolution of the OSS developer's social community to analyze the developer and project data from SourceForge \cite{SourceForge}. According to their ideas, the collaborative social network is formed in global virtual self-organizing teams of software developers working on open source software projects and they developed the self-organizing simulation model. Because the self-organization mechanism is the important property of many social networks, a self-organizing network grows and develops structure without centralized decision making as shown in the real open source software community.

In their model, the developer in the open source software community is an agent. Some agents create a new project (community). The project is the team to develop a software (system). The agent can select only an action representing a real developer's possible daily interaction in the process of developing system. The model has 4 kinds of behaviors: ``Developers can create a project (create a project)'', ``Developers can join into a project (join a project)'',  ``Developers can abandon a project (leave a project)'' or ``Developers can continue the current collaboration (no any actions).'' By such the actions, a new project can be created and be grown, some projects are disappeared. The agent can contact to other agent in the another community and join two or more projects simultaneously. The change in the growing community can be observed through the simulation. The experimental result described in \cite{Madey03} showed the linear relation between the number of projects and developers. 

In our model, we assume that the project has some tasks and the agent has the skill for the developing software. The `task' is the amount of work for the developing software. The `skill' is the amount of works that the agent develops the program for a given time. Our developed agent based simulation model shows some spiral line in the relation between the community and the action of agents. We report the difference of experimental results by using Madey's simulation model and our model under the Source Forge community data.

\section{Open Source Community Data}
\label{sec:OpenSourceCommunityData}
There are some software distribution web site for many open source communities such as GNU project. The Source Forge is one of popular software repositories and has the function of the management system of open source software. The website of Source Forge gives the information of the projects and the developers recorded in the database \cite{SourceForgeData}. 

The data related to project and developer are obtained from the database and are used to investigate the relation between the number of project and developer who joins the project.

\begin{figure}[htbp]
\begin{center}
\includegraphics[scale=0.74]{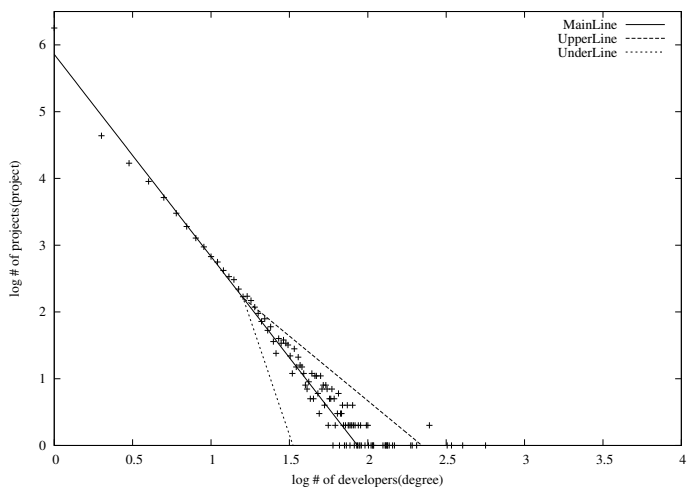}
\caption{The distribution for project and developer in Source Forge}
\label{fig:sf}
\end{center}
\end{figure}

Fig.\ref{fig:sf} shows the logarithm of the number of project and developer at September 2014 (Schema Sf0914). Schema Sf0914 has 73 tables including `User Group.' The table 'User Group' consists of 17 columns including `user\_id' and `group\_id'. We retrieved all data from `sf0914.user\_group' and counted the number of records according to group\_id. As a result, the number of developers to each project is calculated. The horizontal axis is the number of developers and the vertical axis is the number of projects for the numbers of developers. As shown in Fig. \ref{fig:sf}, we can observe that the relation decreased linearly for a certain period and then the line spreads from the certain point like delta. In this paper, we investigate the inflection point and the expanse of the delta and the equation of line is determined according to empirical studies. 

\section{Simulation Model}
\label{sec:SimulationModel}
In our developed simulation system, there are 2 kinds of agents: `major agent' and `minor agent'. The agent has the own function of creating, joining to, or leaving from the project. Because some developers becomes the leadership in the real world of software developing, the simulation model has the 2 kinds of agents behaviors inspired by the interaction behavior of army ant system \cite{Ichimura2014a, Ichimura2014b}.

In our model, a major agent creates the project and a minor agent makes the participation behavior to the project and the leaving behavior from one. A major agent and a minor agent is a project manger and developer, respectively. When some minor agents join the same project, they worked to share the task of software development. The `task' and the `skill' means the numerical values representing the amount of estimated codes for the software development and the agent's coding capability in the periodical time, respectively.

SourceForge can provide open source softwares which are classified into 10 categories such as ``Audio \& Video'', ``Business \& Enterprise'', ``Communications'', ``Development'', ``Home \& Education'', ``Games'', ``Graphics'', ``Science \& Engineering'', ``Security \& Utilities'', ``System Administration'' \cite{SourceForge}. If a major agent creates the project of software, it must determine the project's category. The development work will have two or more types of tasks during developing the software. There are 3 kinds of common types of tasks; `network communication processing', `database processing', and `graphic processing'. The minor agent cannot join the project, if it does not have the skill for the corresponding category. If two or more agents join a project, the task for the developing project is shared to each agent equality.

\subsection{Task and Skill}
A project has one or more tasks in the software development. Let $P_{i}^{k}$ be the amount of task $k$ in the project $i$. $P_{i}^{k}$ takes a value in $[0,2]$. The sum of $P_{i}^{k}$ for all tasks is defined as $\mathbf{P_{i}}=\{P_{i}^{1}, P_{i}^{2}, \cdots, P_{i}^{n} \}$. 

A skill for the agent $j$ in task $k$ is defined as $S_j^k$. The initial value is a given random number in $[0,1]$. The agent works the allocated task with the agent's skill $S_{j}^{k}$. That is, the time to finish the task is calculated by dividing the amount of task of an agent $W^{k}_{ij}$ by the agent skill $S_{j}^{k}$. The working time for each project to which the agent joins is added. The total time is the summation of $T_{j}$ for each agent as follows.
\begin{eqnarray}
T_{j} = \sum_{i} T_{ij} = \sum_{i} \frac{W^{k}_{ij}}{S_{j}^{k}} ,  
\label{eq:totaltime}
\end{eqnarray}
where $W_{ij}^{k}$ is the amount of the task $k$ which is assigned to the agent $j$ in the project $i$. That is, $W_{ij}^{k} = \frac{P_{i}^{k}}{N_{i}}$. $T_{j}$ satisfies the condition $0 \leq T_{j} \leq T_{limit}$. The upper limit of time $T_{limit}$ is $24$, because of 24 hours a day.

\subsection{The behaviors of agent}
\label{sec:agentbehavior}
In \cite{Madey03}, the proposed model has 4 kinds of behaviors: ``Developers can create a project (create a project)'', ``Developers can join into a project (join a project)'',  ``Developers can abandon a project (leave a project)'' or ``Developers can continue the current collaboration (no any actions).'' In this paper, the model has 2 kinds of agents: the major agent and the minor agent. The following actions are implemented by 2 kinds of agents, respectively.

\begin{itemize}
  \item Major agent
  \begin{itemize}
    \item Create a project
  \end{itemize}
  \item Minor agent
  \begin{itemize}
    \item Join a project
    \item Leave a project
    \item Take nothing action
  \end{itemize}
\end{itemize}

A minor agent selects an action from 3 kinds of behaviors per a step and it works according to the selected behavior by the agent's action algorithm.

\subsection{Create a project}
\label{sec:create}
A major agent creates a new project, 
\begin{itemize}
\item when it starts to develop a new software. In this case, a created project is called $New$ one.
\item when the agent creates the sub projects to share the large tasks in the project which is created by the major agent. In this case, a created project is called $Sub$ one.
\end{itemize}

If the major agent $j$ creates a new project, the agent works to follow next behavior's algorithm.

\begin{enumerate}
\item The probabilities $New$ or $Sub$ of creating a project for each case are given appropriate values.
\item If a major agent creates a new project, an agent checks the condition that the probability $New$ equals to or greater than the random value $[0,1]$ by using Monte Carlo method.
\item If a major agent creates a sub project to share the tasks, the major agent determines by Eq.(\ref{eq:SubProbability}) and Eq.(\ref{eq:SubCondition}).
\end{enumerate}
\begin{eqnarray}
Sub^{k}_{i} =  \frac{P^{k}_{i}}{N^{k}_{i}}, 
\label{eq:SubProbability}
\end{eqnarray}
where $Sub^{k}_{i}$ is the average amount of tasks of an minor agent in the developing task $k$ of the project $i$. $P^{k}_{i}$ is the task $k$ assigned to the agent and ${N^{k}_{i}}$ is the number of agents in the project $i$. 
\begin{eqnarray}
\left\{
\begin{array}{l}
IF\;\; Sub^{k}_{i} \geq  Sub_{threshold}\;\; THEN \;CreateProject\\
IF\;\; Sub^{k}_{i} < Sub_{threshold}\;\; THEN \;NotCreate , 
\end{array}
\right.
\label{eq:SubCondition}
\end{eqnarray} 
where $Sub_{threshold}$ is a given value to share the tasks and is a threshold to create a sub project.

The major agent checks the value of $Sub^{k}_{i}$ for each task which should be shared in the point of load balance. Once a sub project is created, some minor agents with skills which is corresponded to the category of developing task will participate to it.

\subsection{Join a project}
\label{sec:Join}
The minor agent $j$ joins the project according to the following process.

\begin{enumerate}
\item The minor agent $j$ selects one project which it has not participated yet.
\item The probability to be joined in the selected project is determined by Eq.(\ref{eq:JoinProbability}).
\item The minor agent determines to join the project by Eq.(\ref{eq:JoinCondition}).
\end{enumerate}
If the estimated time till the end of the task exceeds the threshold $T_{limit}$, the minor agent gives up the joining the project.

\subsubsection{Selection of project with Network growing model}
When the agent selects a project, some strategies to a network growing model are required. In this paper, Barabasi-Albert(BA) model with dynamic fitness\cite{DynamicBA} is adapted to decide the behaviors of agents. BA model is an algorithm for generating random scale-free networks based on the preferential attachment mechanism. Scale-free network is widely used in natural and human-made systems such as social community. In BA model, an edge is connected to the node with high number of degrees in high probability\cite{BAmodel}. BA model with dynamic fitness employs the fitness of node is dynamically changed \cite{DynamicBA}. A degree of software is the number of agents worked in the software development and a minor agent selects a software based on this model. In this model, the linkage to the other agent is implemented by using BA model, but a new node is not added.

The selection probability $S_{i}$ for the software $i$ by a minor agent is defined as follows.
\begin{eqnarray}
S_{i} = \frac{\eta_{i} k_{i}}{\sum \eta_{i} k_{i}}, 
\label{eq:Select}
\end{eqnarray}
where $\eta_{i}$ is the time decreasing function representing the fitness of software $i$ and $k_{i}$ is the degree of software, that is, the number of participating agents. The dynamic fitness function realizes the growing community, the stability and the attenuation of community growing.

\subsubsection{Probability of joining a project}
\label{sec:JoinProbability}
The probability of participation to a project is calculated by the tasks of the project and the skill of the agent $j$. If the agent does not have enough skills for the category of the project $i$, the probability is 0. Otherwise, the probability is calculated as follows.
\begin{eqnarray}
J_{ij} =\frac{\sum_{k=1}^{n}( P_{i}^{k} -\sum_{j=1}S_{j}^{k})}{\sum_{k=1}^{n} P_{i}^{k}}
\label{eq:JoinProbability}
\end{eqnarray}
\begin{eqnarray}
\left\{
\begin{array}{l}
IF\;\; J_{ij}  \geq  J_{threshold}\;\; THEN \;JointProject\\
IF\;\; J_{ij} < J_{threshold}\;\; THEN \;NotJoin, 
\end{array}
\right.
\label{eq:JoinCondition}
\end{eqnarray} 

where $\sum_{k=1}^n( P_{i}^{k} -\sum_{j=1}S_{j}^{k})$ is how tasks of the project have reached the completed work, $J_{ij}$ is calculated by dividing it by the total amount of tasks $\sum_{k=1}^{n} P_{i}^{k}$. If $J_{ij}$ is larger than threshold $J_{threshold}$, a minor agent joins a project.

\subsection{Leave a project}
\label{sec:Leaveproject}
The minor agents can leave from the working project by the following algorithm.

\begin{enumerate}
\item The minor agent $j$ selects one project among all projects to participate.
\item The probability of leaving to participate the project is calculated.
\item A minor agent decides the project to leave by Eq.(\ref{eq:AbandanProbability}).
\end{enumerate}

The probability of leaving a project is calculated as follows.
\begin{eqnarray}
L_{ij} = \sum_{k=1}^n (\frac{P_i^k}{N_i^k} - S_j^k)
\label{eq:AbandanProbability}
\end{eqnarray}
\begin{eqnarray}
\left\{
\begin{array}{l}
IF\;\; L_{ij} \geq  L_{threshold}\;\; THEN \;LeaveProject\\
IF\;\; L_{ij} < L_{threshold}\;\; THEN \;NotLeave , 
\label{eq:AbandanCondition}
\end{array}
\right.
\end{eqnarray}
where $\frac{P_i^k}{N_i^k}$ is the amount of task per an agent, $L_{ij}$ is calculated by subtracting the agent skill from $\frac{P_i^k}{N_i^k}$, and $L_{threshold}$ is a given threshold value. If the load of the agent $j$ ($L_{ij}$) is greater than $L_{threshold}$, the agent $j$ leaves from a project $i$.

In Eq.(\ref{eq:AbandanProbability}), the minor agent can leave from a project with ease, when the amount of the task of the project is high or the skill of agent is low or in both situations.

\section{Simulation Results}
This section describes the effects of the growing community by introducing the concept related to the tasks and the agent skills through the simulation results. In this paper, we prepared the simulation environment with 1,000 major agents and 20,000 minor agents. The threshold values $J_{threshold}$ in Eq.(\ref{eq:JoinCondition})and $L_{threshold}$ in Eq.(\ref{eq:AbandanCondition}) are used to find the optimal set by using the following patterns. The parameters are important to decide the action of the minor agents.
\begin{eqnarray}
\nonumber
(J_{threshold}, L_{threshold}) = (0.0, 0.0), (0.5, 0.0), (1.0, 0.0),\\
\nonumber
(0.0, 0.5) , (0.5, 0.5) , (1.0, 0.5) , (0.0, 1.0), (0.5, 1.0), (1.0, 1.0)
\end{eqnarray}

\begin{figure}[htbp]
\begin{center}
\includegraphics[scale=0.73]{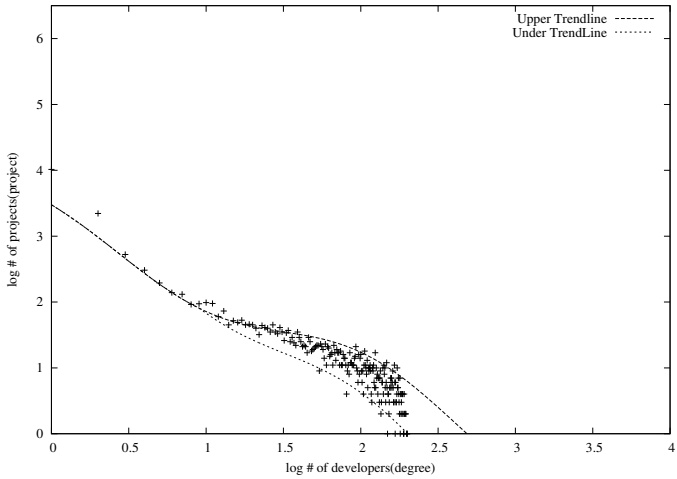}
\caption{The relation between developers and projects with BA model}
 \label{fig:BADynamicDistribution}
\end{center}
\end{figure}

Fig.\ref{fig:BADynamicDistribution} shows the relation of the number of developers and the numbers projects at the 1,000 iterations. The horizontal axis is the number of developers and the vertical axis is the number of projects for the numbers of developers. For 9 kinds of parameter sets, the simulation results as shown in Fig.\ref{fig:BADynamicDistribution} are observed. The relation between the developers and the projects shows the linearity, if the number of developers is less than 1. Otherwise, the relation has the spread in the direction of the number of projects. In case of 1 or more developers, the curve of the relation depicts the spiral shape with polar coordinate. We tried to express the characteristic pattern as the following equations.

\begin{eqnarray}
\left\{
\begin{array}{lll}
r &=& a(\theta) \times \theta\\
a(\theta) &=& 0.07 - 0.07 \times \frac{\theta}{2 \pi}
\label{eq:r}
\end{array}
\right.
\end{eqnarray}

\begin{eqnarray}
\left\{
\begin{array}{lll}
r &=& a(\theta) \times \theta - \delta  \;(\theta_1 \leq \theta)
\label{eq:delta}\\
\delta &=& 1.6 \times \frac{\theta - \theta_1}{2 \pi} , 
\label{eq:delta}\\
\end{array}
\right.
\end{eqnarray}
where $\delta$ means the width in the spread range. $\theta$ is the angle in the polar coordinate and $\theta_1$ is the angle at the appropriate point where the number of developers reaches 1.

The characteristic patten drawn by Eq.(\ref{eq:r}) and Eq.(\ref{eq:delta}) can be seen in 9 kinds of parameter sets.

\begin{figure}[t]
\begin{center}
\includegraphics[scale=0.73]{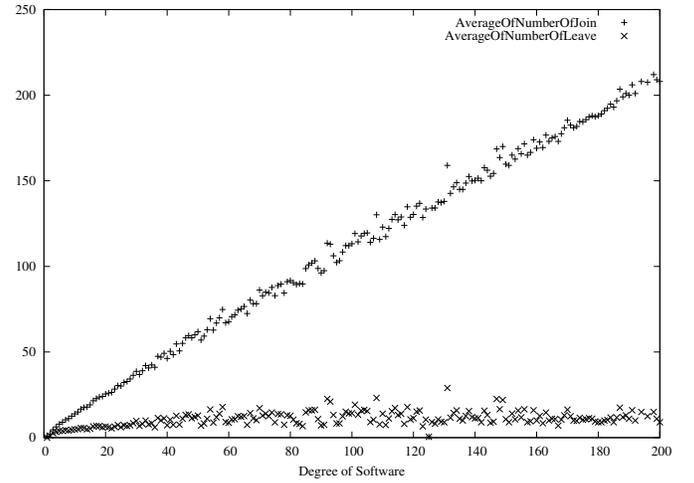}
\caption{The number of the participate agents and the leaving agents}
\label{fig:JoinLeave}
\end{center}
\end{figure}

\begin{figure}[t]
\begin{center}
\includegraphics[scale=0.73]{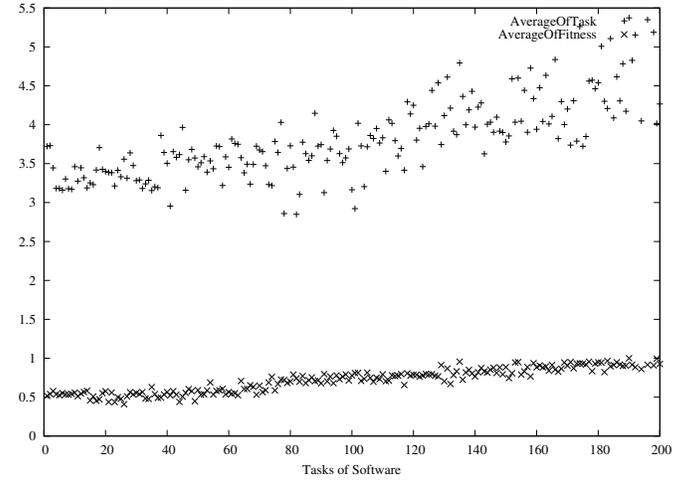}
\caption{The relation between the task and the fitness}
\label{fig:TaskFitness}
\end{center}
\end{figure}

\begin{figure}[htbp]
\begin{center}
\includegraphics[scale=0.73]{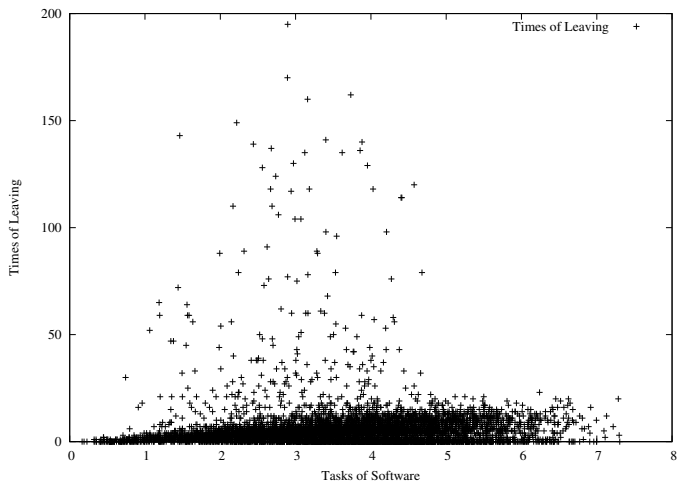}
\caption{The leaving agents for the amount of tasks}
\label{fig:TaskLeaving}
\end{center}
\end{figure}

Moreover, we consider the relation between the behaviors of the agent and the number of agents belonging to the project. Fig.\ref{fig:JoinLeave} shows the number of participating agents and the leaving agents for the tasks in the project. The horizontal axis is the number of tasks of the project and the vertical axis is the number of agents which take any actions. In case of the participating action and the leaving action, as the number of developers becomes large, the distribution approximates linearly. In addition, Fig.\ref{fig:TaskFitness} shows the amount of tasks and the fitness when the task of the projects is created. We can see that the number of tasks in the project is large as the amount of tasks and the value of the fitness becomes high. Because the agent can select the tasks with high fitness easily by Eq.(\ref{eq:Select}) and the agent can participate the project with the large tasks easily by Eq.(\ref{eq:JoinProbability}).

Fig.\ref{fig:TaskLeaving} shows the changing number of leaving times as the amount of task increases. The increase tendency of the leaving agents appeared in the direction of large tasks of the software. Too many tasks in the developing software keep the agents at a distance. Especially, if there are many agents with the low capability in the developing software, that is, the load per agent is high, the leaving agents may increase. The situation may occur within 5 tasks in the software.

\section{Conclusion}
With respect to community construction in software development, we examined with the agent based simulation with representing the relation between the task of project and the skill of the agents. The action that an agent selects software project by BA model with dynamic fitness. We can see that the project in the software with the larger tasks and with the higher fitness are easy selected by many agents. Because the participating probability becomes greater in case of the large task and the larger fitness value.

The solution for the effectiveness of software development will be expected for more complex system development. The altruism behavior \cite{Ichimura2014a,Ichimura2014b}, which is a characteristic behavior of army ant and is a behavior that an army ant sacrifices its own body for the benefit of another ants, will be investigated in the participation behavior and leaving behavior to the projects in software development in future.

\section*{Acknowledgment}
This work was partially supported by JSPS KAKENHI Grant Number 25330366.

\end{document}